\documentclass[twocolumn,showpacs,prl]{revtex4}
\usepackage{amsfonts}
\usepackage{amssymb}
\usepackage{amsmath}
\usepackage{graphicx}
\usepackage{dcolumn}
\usepackage{bm}
\usepackage{soul}
\usepackage{epstopdf}

\begin{document}

\title{Evidence of spin liquid with hard-core bosons in a square lattice}
\author{Y.-H. Chan$^{1,2}$ and L.-M. Duan$^{1,2}$}
\affiliation{$^{1}$Department of Physics, University of Michigan, Ann Arbor, Michigan
48109, USA}
\affiliation{$^{2}$Center for Quantum Information, IIIS, Tsinghua University, Beijing,
China}

\begin{abstract}
We show that laser assisted hopping of hard core bosons in a square optical
lattice can be described by an antiferromagnetic $J_{1}$-$J_{2}$ $XY$ model
with tunable ratio of $J_{2}/J_{1}$. We numerically investigate the phase
diagram of the $J_{1}$-$J_{2}$ $XY$ model using both the tensor network
algorithm for infinite systems and the exact diagonalization for small
clusters and find strong evidence that in the intermediate region around $%
J_{2}/J_{1}\sim 0.5$, there is a spin liquid phase with vanishing
magnetization and valence bond orders, which interconnects the Neel state on
the $J_{2}\ll J_{1}$ side and the stripe antiferromagnetic phase on the $%
J_{2}\gg J_{1}$ side. This finding opens up the possibility of studying the
exotic spin liquid phase in a realistic experimental system using ultracold
atoms in an optical lattice.
\end{abstract}

\maketitle

A spin liquid phase is an exotic state of matter that does not break any
symmetry of the Hamiltonian and has no conventional order even at zero
temperature \cite{1}. A number of microscopic Hamiltonians with frustrated
quantum magnetic interaction could support a spin liquid phase \cite%
{1,2,3,4,5,6}. In particular, very recently, numerical investigations based
on complementary methods have found strong evidence that the
antiferromagnetic $J_{1}$-$J_{2}$ Heisenberg model may have a spin liquid
phase in a square lattice \cite{4,5}. On the experimental side, several
materials are suspected to be in a spin liquid phase at very low temperature
\cite{1}. However, due to complication of physics in these materials, it is
hard to make a direct connection of the prediction from the simplified
microscopic models and the phenomenology observed in real materials \cite{1}%
. Ultracold atoms in an optical lattice provides a clean platform to realize
microscopic models to allow for controlled comparison between theory and
experiments \cite{7,7b}. Proposals have been made to implement the
frustrated magnetic models in an optical lattice \cite{8,9} and various
required configurations of the optical lattices have been realized
experimentally \cite{9}. However, the direct magnetic Heisenberg coupling,
which comes from the higher-order super-exchange interaction, is very weak
under typical experimental conditions \cite{8,10}. It is still very
challenging to reach the extremely low temperature required to observe the
ground state of the magnetic Heisenberg model in an optical lattice.

In this paper, we show strong evidence that a spin liquid phase can emerge
in an antiferromagnetic $J_{1}$-$J_{2}$ $XY$ model in a square lattice. The
calculations are based on two complementary methods: the recently developed
tensor network algorithm applied directly to infinite systems \cite{11,11b}
and the exact diagonalization of small clusters which is combined with the
finite size scaling to infer the phase diagram \cite{ED}. Both methods
suggest that in a small region around $J_{2}/J_{1}\approx 0.5$,
magnetization and valence bond solid orders all vanish, indicting a spin
liquid phase as the ground state. Different from a Heisenberg model, a $XY$\
model can be realized with hard-core bosons in an optical lattice. Through
control of the laser assisted hopping in a square lattice \cite{12}, we
propose a scheme to implement the effective antiferromagnetic couplings for
both the neighboring and the next neighboring sites with a tunable ratio of $%
J_{2}/J_{1}$. In this implementation, both $J_{2}$ and $J_{1}$ are
determined by the hopping rates of the hard-core bosons in an optical
lattice, which is much larger than the conventional super-exchange
interaction for ultracold atoms in the Heisenberg model \cite{8,10}. The
large $J_{1}$-$J_{2}$ couplings open up the possibility to experimentally
realize this model and observe its spin liquid phase based on the
state-of-the-art technology.

\begin{figure}[tbp]
\includegraphics[width=8cm]{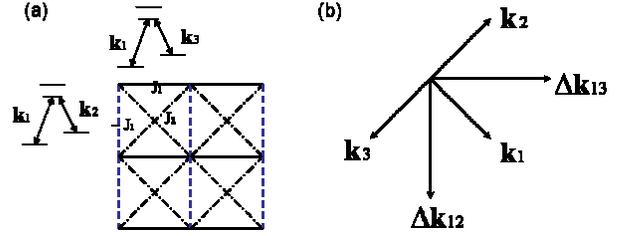}
\caption[Fig. 1 ]{(Color online) (a) Implementation of the $J_1-J_2$ $XY$ model with cold bosons
in a bi-partite square optical lattice, where the $J_2$ coupling is due to the atomic hopping in
the same sub-lattice, and the $J_1$ coupling is induced by the three Raman laser beams (the direct 
$J_1$ hopping of the atoms is turned off by the large potential shift between the two sub-lattices). 
(b) The configuration of the wave-vectors for the three Raman laser beams.}
\end{figure}

The $J_{1}$-$J_{2}$ $XY$ model is represented by the Hamiltonian%
\begin{equation}
H=J_{1}\sum_{\left\langle i,j\right\rangle
}(X_{i}X_{j}+Y_{i}Y_{j})+J_{2}\sum_{\left\langle \langle i,j\right\rangle
\rangle }(X_{i}X_{j}+Y_{i}Y_{j}),
\end{equation}%
where $X,Y$ represent the Pauli operators $\sigma _{x}$ and $\sigma _{y}$, $%
\left\langle i,j\right\rangle $ and $\left\langle \langle i,j\right\rangle
\rangle $ denote respectively the neighboring and the next neighboring sites
in a square lattice as shown in Fig. 1(a). To realize this model with hard
core bosons, we consider ultracold atoms in different hyperfine spins $%
\left\vert a\right\rangle $ and $\left\vert b\right\rangle $ loaded into
alternating square lattices $A$ and $B$ as shown in Fig. 1. This
configuration can be experimentally realized with the spin-dependent lattice
potential \cite{13}. Atoms in spins $\left\vert a\right\rangle $ (or $%
\left\vert b\right\rangle $) freely tunnel in the lattice $A$ (or $B$) with
the hopping rate $t$, however, a direct hopping between the $A,B$ lattices
is forbidden due to the spin-dependent potential shift. Instead, the
inter-lattice hopping is introduced by the laser induced Raman transition as
shown in Fig. 1(a). We use three Raman beams, with wave-vectors $\mathbf{k}%
_{1}$, $\mathbf{k}_{2}$, and $\mathbf{k}_{3}$ and Rabi frequencies $\mathbf{%
\Omega }_{1}$, $\mathbf{\Omega }_{2}$, and $\mathbf{\Omega }_{3}$,
respectively. The directions of the laser beams are shown in Fig. 1(b) with $%
\Delta \mathbf{k}_{12}=\mathbf{k}_{1}-\mathbf{k}_{2}=k_{\Delta }\hat{y}$ and
$\Delta \mathbf{k}_{13}=\mathbf{k}_{1}-\mathbf{k}_{3}=k_{\Delta }\hat{x}$.
The lase induced inter-lattice hopping rates for the neighboring sites are
then given by $t_{x}=\int w^{\ast }\left( x_{i},y_{i}\right) \mathbf{\Omega }%
_{1}^{\ast }\mathbf{\Omega }_{3}e^{ik_{\Delta }x}w\left(
x_{i+1},y_{i}\right) dxdy$, and $t_{y}=\int w^{\ast }\left(
x_{i},y_{i}\right) \mathbf{\Omega }_{1}^{\ast }\mathbf{\Omega }%
_{2}e^{ik_{\Delta }y}w\left( x_{i},y_{i+1}\right) dxdy$, for the hopping
along the $x,y$ directions, respectively. Assume $\mathbf{\Omega }_{3}=-%
\mathbf{\Omega }_{2}$ and the Wannier function $w\left( x_{i},y_{i}\right) $
symmetric along the $x,y$ directions, we have $t_{x}=-t_{y}=t^{\prime }$ (we
can always choose $t^{\prime }>0$ by setting an appropriate relative phase
between $\mathbf{\Omega }_{1}$ and $\mathbf{\Omega }_{3}$). If the on-site
atomic repulsion $U$ satisfies $U\gg t,t^{\prime }$, we have the hard-core
constraint with at most one boson per site. The hard-core bosons in this
square lattice are then described by the Hamiltonian
\begin{equation}
H=t^{\prime }\sum_{\left\langle i,j\right\rangle _{x}}a_{i}^{\dagger
}b_{j}-t^{\prime }\sum_{\left\langle i,j\right\rangle _{y}}a_{i}^{\dagger
}b_{j}-t\sum_{\left\langle \langle i,j\right\rangle \rangle }(a_{i}^{\dagger
}a_{j}+b_{i}^{\dagger }b_{j})+H.c.
\end{equation}%
The hard core bosons $a_{i},b_{j}$ satisfy the same commutators as the Pauli
operators $\sigma _{i}^{-}$, $\sigma _{j}^{-}$, so with the mapping $%
a_{i}\longrightarrow \sigma _{i}^{-}$ and $b_{j}\longrightarrow \sigma
_{j}^{-}$ for the odd numbers of rows, and $a_{i}\longrightarrow -\sigma
_{i}^{-}$ and $b_{j}\longrightarrow -\sigma _{j}^{-}$ for the even numbers
of rows, the Hamiltonian (2) is mapped to the $J_{1}$-$J_{2}$ $XY$ model in
Eq. (1) with $J_{1}=t^{\prime }/2>0$ and $J_{2}=t/2>0$. Apparently, the
ratio $J_{2}/J_{1}$ is tunable by changing the magnitude of the Rabi
frequencies $\mathbf{\Omega }_{1}^{\ast }\mathbf{\Omega }_{3}$.

In the following, we calculate the phase diagram of the Hamiltonian (1) as a
function of the dimensionless parameter $J_{2}/J_{1}$ ($J_{1}$ is taken as
the energy unit). In the limit $J_{2}/J_{1}\ll 1$, the $J_{1}$ term
dominates and the ground state is magnetized with a Neel order at the
momentum $k=(\pi ,\pi )$. In the opposite limit $J_{2}/J_{1}\gg 1$, the
ground state has a stripe magnetic order at the momentum $(\pi ,0)$ or $%
(0,\pi )$, which minimizes the energy of the $J_{2}$ term. In the
intermediate region with $J_{2}/J_{1}\sim 0.5,$ the Hamiltonian is highly
frustrated with competing interaction terms. Our main purpose is to find out
the phase diagram in this region through controlled numerical simulations.

Our numerical simulations are based on two complimentary methods: exact
diagonalization (ED) for small clusters \cite{ED} and tensor network
simulation for infinite systems \cite{11,11b}. The ED method is limited by
the cluster size, and we use extrapolation based on the finite-size scaling
to infer the phase diagram for the infinite system. The tensor network
algorithm is an recently developed simulation method inspired by quantum
information theory \cite{11}. It can be considered as an extension of the
density matrix renormalization group (DMRG) method to the two dimensional
case, replacing the matrix product state in the DMRG method\ with the tensor
network state that better matches the geometry of the underlying lattice
\cite{11}. We use a particular version of the tensor network algorithms, the
infinite projected entangled pair states (iPEPS) method \cite{11b}, which
applies directly to infinite systems using the translational symmetry. To
take into account the ordered states for the Hamiltonian (1) that
spontaneously break the translational symmetry, in our simulation we take a
unit cell (typically $2\times 2$ and $4\times 4$) that is large enough to
incorporate the relevant symmetry breaking orders \cite{14}. We apply
imaginary time evolution to reach the ground state of the Hamiltonian. To
avoid being stuck in a metastable state, we take a number of random initial
states for the imaginary time evolution and pick up the ground state as the
one which has the minimum energy over all the trials. The accuracy of the
iPEPS simulation depends on the internal dimension $D$ of the tensor network
state. The simulation time scales up very rapidly with the dimension $D$,
which limits $D$ to a small value in practice. We typically take $D$ between
$4$ to $6$ in our simulation.

Figure 2 shows the major result from the iPEPS simulation. First, we look at
the average magnetization $m_{s}=\left( 1/N_{s}\right) \sum_{i}\sqrt{%
X_{i}{}^{2}+Y_{i}{}^{2}+Z_{i}{}^{2}}$ as a function of $J_{2}/J_{1}$, where
the average is taken over the $N_{s}$ sites in the unit cell. The
calculation shows that for small or large $J_{2}/J_{1}$, the ground states
are magnetic (with the Neel or the stripe order, respectively), which is
consistent with our intuitive picture. In the intermediate region with $%
0.46\precsim J_{2}/J_{1}\precsim 0.54$, there is a sudden drop of all the
magnetic orders to a tiny value. Although the iPEPS method under a small
dimension $D$ could be biased toward a less entangled state, which is
typically an ordered state, it would not be baised toward a disordered spin
liquid state. So, when we see a big sudden drop of the magnetic orders from
the simulation, it must be a real effect, strongly indicating there is a new
phase in the intermediate region with vanishing magnetic orders. The
remaining small $m_{s}$ may be due to the finite dimension $D$ and should
vanish when $D$ is scaled up.

To figure out the property of the phase in the intermediate region, we
further check different kinds of valence bond solid orders. We calculate all
the neighboring valence bonds $\left\langle \mathbf{\sigma }_{i}\cdot
\mathbf{\sigma }_{j}\right\rangle $ in the unit cell and the result is shown
in Fig. 2. For a valence bond solid state, the spatial symmetry should be
spontaneously broken for the valence-bond distribution. Figure 2 shows that
in the entire region of $J_{2}/J_{1}$, the valence bond distribution has the
same symmetry as the underlying Hamiltonian, which indicates that the ground
state of the Hamiltonian (1) has no valence bond solid orders. Together with
the above calculation of the magnetic orders, this suggests that the
Hamiltonian (1) has a spin liquid phase with no orders in the intermediate
region with $0.46\precsim J_{2}/J_{1}\precsim 0.54$. This spin liquid phase
seems to have the same feature as the $Z2$ spin liquid in the intermediate
coupling region of the $J_{1}$-$J_{2}$ Heisenberg model found in the recent
numerical simulation \cite{4,5}.

\begin{figure}[tbp]
\includegraphics[width=9cm]{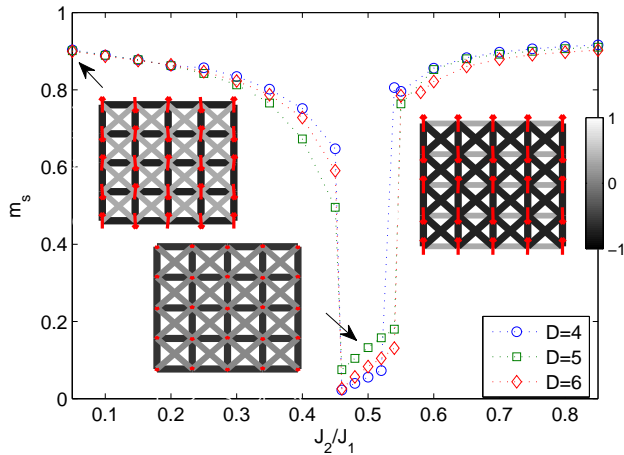}
\caption[Fig. 2 ]{(Color online) Average magnetization $m_s$ as a function of  $J_2/J_1$. 
The insets show the spin configuration and the valence bond distribution $\left\langle \mathbf{\sigma }_{i}\cdot
\mathbf{\sigma }_{j}\right\rangle $
at $J_2/J_1=0$, $0.5$, and $0.9$ obtained with the iPEPS on a $%
4\times4$ unit cell with $D=6$. The width and color of the bonds are scaled
such that the negative energy is shown by thicker bond with darker color and
the positive energy is shown by thinner bond with lighter color and the
length of the spin is proportional to its magnetic moment $m_s$.}
\end{figure}

\begin{figure}[tbp]
\includegraphics[width=8cm]{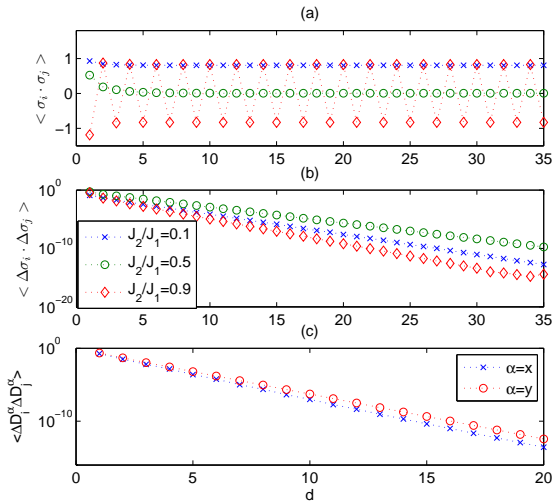}
\caption[Fig. 3 ]{(Color online) (a) Spin-spin correlation $%
\left\langle \mathbf{\sigma }_{i}\cdot \mathbf{\sigma }_{j}\right\rangle $ as a function of distance $%
d$ along the diagonal direction at $J_2/J_1=0.1$ (cross), $0.5$ (
circle) and $0.9$ (open diamond). (b) Semi-log plot of spin-spin correlation $%
\left\langle \Delta \mathbf{\sigma }_{i}\cdot \Delta \mathbf{\sigma }_{j}\right\rangle $ after
subtracting the local averages. (c) Semi-log plot of
dimer-dimer correlation $\langle
\Delta D_{i}^{\alpha}\Delta D_{j}^{\alpha}\rangle$ ($\alpha=x,y$) 
as a function of distance $d$ along the diagonal direction at $J_2/J_1=0.5$.}
\end{figure}

To further confirm this picture, we calculate the long-range spin
correlation and dimer correlation with the iPEPS method and the result is
shown in Fig. 3 for $J_{2}/J_{1}=0.1,0.5$ and $0.9$. The spin correlation $%
\left\langle \mathbf{\sigma }_{i}\cdot \mathbf{\sigma }_{j}\right\rangle $
is calculated along the diagonal direction $\hat{x}+\hat{y}$ . Both the N\'{e%
}el and the stripe phases have long-range correlations, with constant or
staggered values along the diagonal direction. The intermediate phase has an
exponentially decaying spin-spin correlation, which is in agreement with the
behavior of the $Z2$ spin liquid phase with a finite spin gap \cite{1,4}.
The dimer operator $D_{i}^{\alpha }$ is defined by $D_{i}^{\alpha }=\mathbf{%
\sigma }_{i}\cdot \mathbf{\sigma }_{i+\alpha }$ for the bond $(i,i+\alpha )$%
, where $\alpha =\hat{x}$ or $\hat{y}$ denote the orientation of the dimer.
In Fig. 3(c), we show the dimer-dimer correlations $\langle
\Delta D_{i}^{x}\Delta D_{j}^{x}\rangle $ and $\langle \Delta D_{i}^{y}\Delta D_{j}^{y}\rangle $ at $%
J_{2}/J_{1}=0.5$ along the diagonal direction. The correlations are
exponentially decaying with distance, in agreement with a spin liquid phase
with no dimer orders.

In the following, we present study of the Hamiltonian (1)\ with the
complementary ED\ method, which provides further evidence for a spin liquid
phase in the intermediate region. To be consistent with the periodic
boundary condition required for the finite size scaling and to incorporate
the momentum $k=(\pi ,\pi )$ responsible for the Neel order, the size of the
clusters for the ED is taken to $16$, $20$ and $32$ sites. From the spin
correlation $\left\langle \mathbf{\sigma }_{i}\cdot \mathbf{\sigma }%
_{j}\right\rangle $, we calculate the corresponding static structure factor $%
m_{s}^{2}(\mathbf{k},N)=(1/N)\sum_{ij}e^{i\mathbf{k}\cdot (\mathbf{r}_{i}-%
\mathbf{r}_{j})}\left\langle \Delta \mathbf{\sigma }_{i}\cdot \Delta \mathbf{%
\sigma }_{j}\right\rangle $, where $N$ is the size of the cluster and $%
\Delta \mathbf{\sigma }_{i}\equiv \mathbf{\sigma }_{i}-\left\langle \mathbf{%
\sigma }_{i}\right\rangle $. The Neel order and the stripe order correspond
to peaks at $\mathbf{k}=(\pi ,\pi )$ and $(\pi ,0)$, respectively.
Finite-size clusters always have non-zero order parameters, and one needs to
do finite size scaling, with a simple scaling formula $m_{s}^{2}(\mathbf{k}%
,N)=m_{s}^{2}(\mathbf{k},\infty )+a/\sqrt{N}$ ($\sqrt{N}$ corresponds to the
linear size), to infer the value of $m_{s}^{2}(\mathbf{k},\infty )$ for the
infinite system. In Fig. 4, we show the finite size scaling for $m_{s}^{2}(%
\mathbf{k},N)$ at $J_{2}/J_{1}=0,0.5$, and $0.9$ in three different regions.
The results are consistent with the findings from~iPEPS method, i.e., there
is a stripe order with $\mathbf{k}=(\pi ,0)$ at $J_{2}/J_{1}=0.9$ and a Neel
order with $\mathbf{k}=(\pi ,\pi )$ at $J_{2}/J_{1}=0$. At $J_{2}/J_{1}=0.5$%
, the finite-size scaling indicates a vanishing stripe order. However, at $%
\mathbf{k}=(\pi ,\pi )$, the data become non-monotonic with $N$ due to the
shape of the cluster and the finite-size scaling becomes inconclusive in
this case. The non-monotonic shape effect has also been observed in ED for
the $J_{1}$-$J_{2}$ Heisenberg model \cite{ED}.

To check for possible valence bond solid orders from ED, we similarly
calculate the structure factors $m_{d}^{2}(\mathbf{k},N)=(1/N)\sum_{ij}e^{i%
\mathbf{k}\cdot (\mathbf{r}_{i}-\mathbf{r}_{j})}\langle \Delta
D_{i}^{x}\Delta D_{j}^{x}\rangle $ and $m_{p}^{2}(\mathbf{k}%
,N)=(1/N)\sum_{ij}e^{i\mathbf{k}\cdot (\mathbf{r}_{i}-\mathbf{r}%
_{j})}(\langle \Delta P_{i}\Delta P_{j}\rangle $, corresponding respectively
to the dimer order $D_{i}^{x}$ and the plaquette order $%
P_{i}=(Q_{i}+Q_{i}^{-1})/2$, where $Q_{i}$ ($Q_{i}^{-1}$) is the clockwise
(anticlockwise) cyclic permutation operator on the plaquette $i$ with its
explicit (lengthy) expression given in \cite{14,15}. The rotational symmetry
is always preserved at finite size, so we only need to check one component
of the dimer order, say $D_{i}^{x}$. At finite size, the structure factors
peak at $\mathbf{k}=(\pi ,0)$ for the dimer order $D_{i}^{x}$ and at $%
\mathbf{k}=(\pi ,\pi )$ for the plaquette order $P_{i}$, however, an
extrapolation to the infinite system at these momenta as shown in Fig. 5
indicates vanishing dimer and plaquette orders in all three regions of $%
J_{2}/J_{1}$. This result, again, is in agreement with the finding from the
iPEPS calculation.

\begin{figure}[tbp]
\includegraphics[width=8cm]{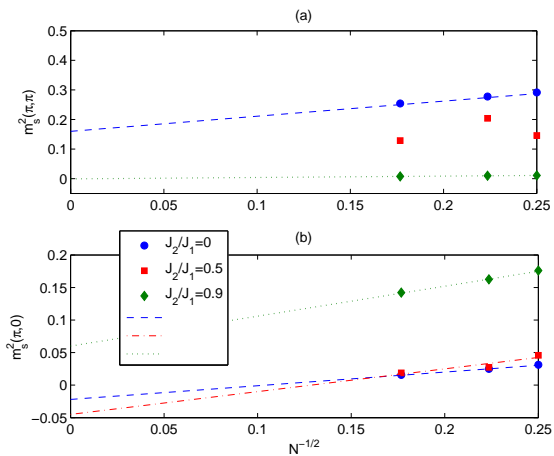}
\caption[Fig. 4 ]{(Color online) Finite size scaling of the magnetic
order parameter at (a) $\mathbf{k}=(\protect\pi,\protect\pi)$ and (b) $%
\mathbf{k}=(\protect\pi,0)$ at $%
J_2/J_1=0$ (dot), $0.5$ (square), and $0.9$ (diamond). }
\end{figure}

\begin{figure}[tbp]
\includegraphics[width=8cm]{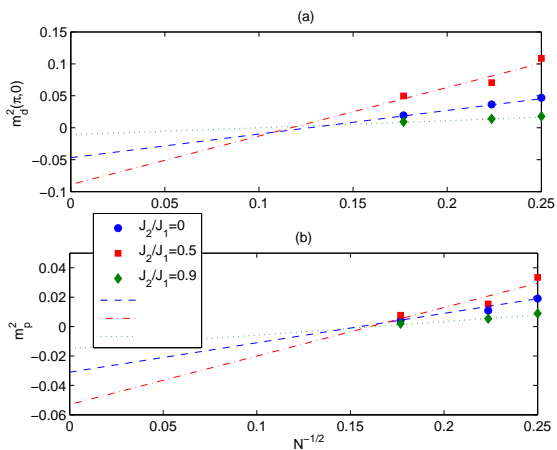}
\caption[Fig. 5 ]{(Color online) Finite size scaling of (a) the dimer
order parameter at $\mathbf{k}=(\protect\pi,0)$ and (b) the plaquette
order parameter at $\mathbf{k}=(\protect\pi,\pi)$ at $%
J_2/J_1=0$ (dot), $0.5$ (square), and $0.9$ (diamond).}
\end{figure}

Before concluding the paper, we briefly discuss the experimental signature
of the three different phases for the Hamiltonian (1) in the implementation
with hard-core bosons. The Neel ordered state and the strip phase correspond
to Bose-Einstein condensates at the momenta $\mathbf{k}=(\pi ,\pi )$ and $%
\mathbf{k}=(\pi ,0)$, respectively. The standard time-of-flight imaging
measurement can then reveal the condensate peak at these nontrivial momentum
points \cite{7b}. The spin liquid phase, on the other hand, would not show
any condensation peaks due to lack of magnetic orders. Furthermore, it has a
spin gap which implies a charge gap in implementation with hard-core bosons.
We therefore expect to see an incompressible phase at half filling, which is
different from the Mott insulator state at the integer filling. It is also
be distinguished from a charge density wave state since the density
distribution in this case is still homogeneous without any solid order.

In summary, we have proposed an experimentally feasible scheme to implement
the $J_{1}$-$J_{2}$ $XY$ model with ultracold hard-core bosons in a square
optical lattice. Through detailed numerical simulation of this model using
two complementary methods, we find strong evidence that this model has a
spin liquid phase in the intermediate region of $J_{2}/J_{1}$. The proposed
experimental implementation, with tunable ratio of $J_{2}/J_{1}$, opens up a
realistic possibility to look for the long-pursued spin liquid phase in a
well controlled Hamiltonian model.

We thank Hsiang-Hsuan Hung and Hong-Chen Jiang for helpful discussion. This
work was supported by the NBRPC (973 Program) 2011CBA00300 (2011CBA00302),
the DARPA OLE program, the IARPA MUSIQC program, the ARO and the AFOSR MURI
program.

\end{document}